\documentstyle[eqsecnum,prd,aps,epsfig]{revtex}
\begin{document}
%
%
\def\ov{\over}
\def\l{\left}
\def\r{\right}
\def\be{\begin{equation}}
\def\ee{\end{equation}}
\draft
\title{Spherical neutron star collapse toward a black hole in
tensor-scalar theory of gravity}
\author{J\'er\^ome Novak}
\address{D\'epartement d'Astrophysique Relativiste et de Cosmologie \\
  UPR 176 du C.N.R.S., Observatoire de Paris, \\
  F-92195 Meudon Cedex, France}
\date{January 12, 1998}
\maketitle

\begin{abstract}
Complete tensor-scalar and hydrodynamic equations are presented and
integrated, for a self-gravitating perfect fluid. The initial
conditions describe unstable-equilibrium neutron star
configuration, with a polytropic equation of state. They are necessary
in order to follow the gravitational collapse (including full
hydrodynamics) of this star toward a black hole and to study the
resulting scalar gravitational wave. The amplitude of this wave, as
well as the radiated energy dramatically increase above some critical
value of the parameter of the coupling function, due to the spontaneous
scalarization, an effect not present in Brans-Dicke theory. In most
cases, the pressure of the collapsing fluid does not have a
significant impact on the resulting signal.  
These kind of sources are not likely to be observed by future laser
interferometric detectors (such as VIRGO or LIGO) of gravitational
waves, if they are located at more than a few $100$ kpc. However,
spontaneous scalarization could be constrained if such a
gravitational collapse is detected by its quadrupolar gravitational
signal, since this latter is quite lower than the monopolar one.

\end{abstract}

\pacs{PACS number(s): 04.30.Db, 04.50.+h, 04.25.Dm, 02.70.Hm}

\section{Introduction}
\label{s:intro}

In order to test general relativity, one has to compare it to other, 
alternate theories of gravitation. {\em Tensor-scalar} theories, 
in which gravity is described by a spin-2 field combined with one or several
spin-0 fields, are not only alternate theories, but ``generalize general 
relativity'' (see \cite{DEF92} ); meaning that general relativity is obtained
in them by setting all scalar fields to zero. Several such theories have been
developed, from Fierz \cite{FIE56}, Jordan \cite{JOR59}, Brans and Dicke
\cite{BD61} to Bergman \cite{BER68}, Nordtvedt \cite{NOR70}, Wagoner
\cite{WAG70} and, more recently, Damour and Esposito-Far\`ese \cite{DEF92}.
In all these theories, the spin-0 and spin-2 fields ($\varphi$ and
 $g_{\mu \nu}$) are coupled to matter via an effective metric tensor
 $\tilde{g}_{\mu \nu}=a^2(\varphi)g_{\mu \nu}$. The Jordan-Fierz-Brans-Dicke
theory has only one free parameter $\omega$, whereas for Bergman, Nordtvedt and
Wagoner the parameter is a function $\omega(\varphi)$. Damour and 
Esposito-Far\`ese considered an arbitrary number of scalar fields, coupled
one to the other. All these theories are motivated by, mainly, two 
theoretical reasons:
\begin{enumerate}
\item they represent the low-energy limit of superstring theories (\cite{CFM85}
 and \cite{DP94})
\item they give rise to new ``extended'' inflationary models \cite{SA90}
\end{enumerate}

Since in Brans-Dicke theory $\log(a(\varphi))$ is a linear function of
 $\varphi$, solar system experiments (weak field) are sufficient to
constrain the theory, even in strong fields. Nevertheless, a more
general theory, in which $\log(a(\varphi))$ is a parabolic function (depending
on two parameters), shows non-perturbative effects in strong field 
\cite{DEF93}, described as ``spontaneous scalarization'' in \cite{DEF96}.
Thus, when describing neutron stars, general relativity
and tensor-scalar theory can give significant differences for their masses,
radii and gravitational field, whereas the difference can still be negligible
in our solar system data. As a consequence, 
weak-field experiments {\em cannot} give much information on
strong-field regime, and one needs to test this strong-field regime by other
means:
\begin{itemize}
\item[\_] by looking for the orbital decay of binary-pulsar
systems. This has been 
done by Damour and Esposito-Far\`ese (\cite{DEF92} and \cite{DEF96}), who 
constrained the parameter space of the coupling function,
\item[\_] by looking for monopolar gravitational radiation from
collapsing compact  sources  which could be detected by
the laser interferometric gravitational wave observatories (such as VIRGO
\cite{VIRGO} and LIGO \cite{LIGO}).
\end{itemize} 

This latter method requires that the signal be known and that the
observed (or unobserved) amplitude be related to the coupling
function parameters. Such kind of computations have already been 
performed by various groups, but they all considered only an
Oppenheimer-Snyder 
collapse (i.e. ``dust'' matter, with no pressure), either in Brans-Dicke
theory (\cite{SNN94} and \cite{SST95}) or by doing some Taylor expansion of
the coupling function \cite{HCN97}. In this latter, the parameter
space of the coupling function was restricted to the part where
non-perturbative strong-fields effects do not happen. The aim of this
paper is to present the 
results of computations of a spherically symmetric collapse, of a
neutron star toward a black hole \footnote{numerically, the black hole
is never obtained, but the monopolar gravitational waves, far from the
source, behave as if the black hole had formed (see \ref{s:scagw})},
with one scalar field and an arbitrary 
coupling function. All the hydrodynamics and field equations are treated
with no approximation in order to get the monopolar gravitational wave
form and amplitude. Moreover, including the equation of state allows
to start the collapse with quite a realistic neutron star
configuration and thus, spurious waves signals are avoided (see
\ref{s:compar}). 

The paper in organized as follows. Section~\ref{s:equa} describes the
evolution equations for the star; section~\ref{s:resu} gives the
numerical results: initial-value models (\ref{s:stat}), collapse and
resulting wave signal (\ref{s:scagw}), comparison with previous
works (\ref{s:compar}) and exploration of the parameter space
(\ref{s:expl}); finally section~\ref{s:conc} gives some concluding
remarks.

\section{Field and hydrodynamic equations}
\label{s:equa}
\subsection{General equations}

As it has been stated before, the most general theory containing a spin-2
field and one (massless) spin-0 field contains one arbitrary coupling
function $a(\varphi)$. The action is given by
\be
S=(16\pi G_*)^{-1}\int d^4x\,\sqrt{-g_*}(R_*-2g_*^{\mu \nu}\partial_\mu 
\varphi \partial_\nu \varphi) + S_m[\Psi_m,a^2(\varphi)g^*_{\mu \nu}]
\ee
where all quantities with asterisks are related to the
``Einstein metric'' $g^*_{\mu \nu}$: $G_*$ is the bare gravitational
 coupling constant,
 $R_*=g_*^{\mu \nu}R^*_{\mu \nu}$ the curvature scalar for this metric and
 $g_* = \det(g^\mu_{*\nu})$. The term $S_m$ denotes the action of the 
matter,
represented by the fields $\Psi_m$, which is coupled to the ``Jordan-Fierz''
metric $\tilde{g}_{\mu \nu}=a^2(\varphi)g^*_{\mu \nu}$; all quantities with
a tilde are related to this metric. That means that all non-gravitational
experiments measure this metric, although the field equations of the theory are
better formulated in the Einstein metric. The indices of Einstein frame
quantities are moved through Einstein metric, whereas those of Jordan-Fierz
quantities are moved through Jordan-Fierz one. By varying $S$, one obtains
\begin{eqnarray}
 && R^*_{\mu \nu} - {1\ov 2}g^*_{\mu \nu}R^*  = 2\partial_\mu \varphi
\partial_\nu \varphi - g^*_{\mu \nu}g_*^{\rho \sigma} \partial_\rho
\varphi \partial_\sigma \varphi + {8\pi G_*\ov c^4} T^*_{\mu \nu} 
\label{e:tscal}\\
 && \Box_{g_*} \varphi = -{4\pi G_* \ov c^4} \alpha(\varphi)T_* \label{e:ondsc}
\end{eqnarray}
where
\begin{eqnarray}
&& T_*^{\mu \nu} = {2\ov \sqrt{-g_*}}{\delta S_m \ov \delta g^*_{\mu \nu}} \\
&&\alpha(\varphi) = {\partial \ln a(\varphi)\ov \partial \varphi}
\end{eqnarray}
and $\Box_{g_*}=g_*^{\mu \nu}\nabla^*_\mu\nabla^*_\nu$ is the Laplace-Beltrami
operator of $g^*_{\mu \nu}$, $\nabla^*_\mu$ denoting the Levi-Civita
connection of $g^*_{\mu \nu}$. One can see that $\alpha(\varphi)$ is the basic,
field-dependent coupling function between matter and scalar field. General
relativity is obtained for $\alpha(\varphi)\to 0$.

 The physical stress-energy tensor
$\tilde{T}^{\mu \nu} = 2(-\tilde{g})^{-1/2}\, \delta S_m / \delta 
\tilde{g}_{\mu \nu}$ is related to the Einstein-frame one by
\be
T^\mu_{*\nu} = a^4(\varphi) \tilde{T}_\nu^\mu
\ee
The equations of motion are given by the stress-energy balance equation,
written in the Jordan-Fierz frame
\begin{eqnarray}
&& \tilde{\nabla}_\nu\tilde{T}_\mu^\nu = 0  \\
\text{and in the Einstein frame } && \nabla^*_\nu T_*^{\mu \nu}=
\alpha(\varphi)T^*\nabla_*^\mu\varphi \label{e:consen}
\end{eqnarray}
Finally, let us call $\varphi_0$ the cosmological value of the scalar field,
which enters the theory as the boundary condition on the scalar field at 
spatial infinity.

\subsection{Coordinates and variables}

The present calculations have essentially been done by generalizing a
previous work by Gourgoulhon \cite{GOU91} to tensor-scalar
theory. Therefore, only a very brief presentation of coordinate and
variable choice will be given here. 
The Einstein-scalar equations have been decomposed in the $3+1$ formalism
\cite{ADM62} onto a family of spacelike hypersurfaces $\Sigma_t$ labeled by
the real index $t$ called the {\em coordinate time}.
The {\em polar time slicing} has been
chosen in order to have good singularity avoidance (see e.g. \cite{PIR83} for
discussion). On each hypersurface $\Sigma_t$ the {\em radial gauge} has been
chosen with spherical-like coordinates $(r,\theta,\phi)$, since the considered
problem is spherically symmetric. All these assumptions (spherical
symmetry, radial gauge and polar slicing, called RGPS) imply that the metric
 $g_{\mu\nu}^*=a^{-2}(\varphi)\tilde{g}_{\mu\nu}$
(which verifies Einstein-like equations (\ref{e:tscal}), with an extra-term)
takes the diagonal form:
\be
ds^2 = -N^2(r,t)dt^2 + A^2(r,t)dr^2 + r^2(d\theta^2 +
 \sin^2\theta d\phi^2)
\ee
where $N(r,t)$ is called the {\em lapse function}. The metric 
$g^*_{\mu\nu}$ will
often be described by the three functions $\nu(r,t)$, $m(r,t)$ and
 $\zeta(r,t)$ defined by
\begin{eqnarray}
N(r,t) = \exp(\nu(r,t)) \\
A(r,t) = \left(1-{2m(r,t)\ov r}\right)^{-1/2} \label{e:defa}\\
\text{ and } \zeta(r,t) = \ln \l({N\ov A}\r)
\end{eqnarray}
All coordinates are expressed in the Einstein-frame, and asterisks are
 omitted. However, ``physical'' quantities will often be written in the
Fierz metric and noted with a tilde.

In this work, neutron stars are modeled as self-gravitating perfect fluids.
They can be considered to be
made of degenerate matter at equilibrium, the equation of state being
temperature independent (cold  matter). This does not hold only soon
after their formation. The stress-energy tensor writes
\be
\tilde{T}_{\mu\nu} = (\tilde{e}+\tilde{p})\tilde{u}_\mu \tilde{u}_\nu 
+ \tilde{p} \tilde{g}_{\mu\nu} \label{e:flparf}
\ee
where $\tilde{u}_\mu$ is the 4-velocity of the fluid, $\tilde{e}$ is the total 
energy density
(including rest mass) in the fluid frame and $\tilde{p}$ is the pressure. 
The relation to its Einstein-frame counterpart is $T^\mu_\nu=a^4(\varphi)
 \tilde{T}^\mu_\nu$. The
description of the fluid is completed by an equation of state
\be
\tilde{e} = \tilde{e}(\tilde{n}_B)
\ee
with $\tilde{n}_B$ being the baryonic density in the fluid frame. One then 
deduces the
pressure as a function of $\tilde{n}_B$. Let $\Gamma=\tilde{N}
\tilde{u}^0$ be the
Lorentz factor connecting the fluid frame and $\Sigma_t$ hypersurface frame,
calling
\be
\tilde{E} = - \tilde{T}^0_0 \text{ one gets } \tilde{E} = 
\Gamma^2(\tilde{e}+\tilde{p}) - \tilde{p}
\ee
The fluid baryonic number is represented by the {\em coordinate baryonic
density}
\be
\tilde{D} = {\text{number of baryons in }\delta V \ov \delta V} =
 A\Gamma \tilde{n}_B \label{e:defd} 
\ee
where $\delta V=r^2\sin \theta dr d\theta d\phi$ is the element of the
coordinate 3-volume on a given $\Sigma_t$, defined as the set of
points whose coordinates are between $r$ and $r+dr$, $\theta$ and
 $\theta+d\theta$, $\phi$ and $\phi+d\phi$. The
fluid motion is described by the following variables
\begin{eqnarray}
V = {dr\ov dt} = {u^r \ov u^0} \text{ (coordinate velocity)} \\
U = { \text{proper distance travelled on } \Sigma_t \ov \text{elapsed
proper time on }\Sigma_t} = {A \ov N} V
\end{eqnarray}
one then has $\Gamma=(1-U^2)^{-1/2}$ and deduces the components of 
$\tilde{T}_{\mu
\nu}$ given by (\ref{e:flparf}).
The fluid {\em log-enthalpy} is also introduced, defined as:
\be
H = \ln\left({\tilde{e}+ \tilde{p} \ov \tilde{n}_B m_B c^2}\right)
\label{e:defenth}
\ee
and, finally, three ``scalar-field'' variables:
\begin{eqnarray}
\eta &=& {1\ov A}{\partial \varphi \ov \partial r} \\
\psi &=& {1\ov N}{\partial \varphi \ov \partial t} \\
\Xi &=& \eta ^2 + \psi ^2 
\end{eqnarray}

\subsection{Tensor-scalar field equations}

Spherical symmetry helps to obtain gravitational field equations; we followed
the procedure described by Gourgoulhon \cite{GOU91}, projecting eqs.~(\ref
 {e:tscal}--\ref{e:ondsc}) on the 3-surfaces $\Sigma_t$ and along their
normal. Hereafter, we use the following notation: 
 $$ q_\pi = {8\pi G_* \ov c^4} $$ 
 
The tensor Einstein-like equations~(\ref{e:tscal}) turn then  into:
\begin{itemize}
\item [\_] one {\em Hamiltonian constraint equation} 
\be
{\partial m \ov \partial r} = r^2 {c^2 \ov 2G_*} \l(\Xi + q_\pi a^4(\varphi)
\tilde{E}\r) \label{e:dmdr}
\ee
\item [\_] three {\em momentum constraint equations} which reduce to only one
non-vanishing
\be 
{\partial m \ov \partial t} = r^2{c^2 \ov 2G_*} \l[ 2{N\ov A}\psi \eta
-q_\pi a^4(\varphi)(\tilde{E}+\tilde{p}) V\r] \label{e:dmdt}
\ee
\item[\_] six  {\em Einstein dynamical equations} which here reduce to two
non-vanishing; one is degenerate, only giving a condition on the lapse
function
\be
{\partial \nu \ov \partial r} = {q_\pi A^2\ov 2} \l[{m c^2\ov 4\pi r^2}+
 a^4(\varphi)r(\tilde{p}+U^2(\tilde{E}+ \tilde{p})) +r \Xi \r] 
\label{e:dnudr}
\ee
The other one will not be used in this work.
\end{itemize}

Writing the scalar-field wave equation (\ref{e:ondsc}) with our variables gives
\be
{\partial^2 \varphi \ov \partial t^2} = e^{2\zeta} \l(\Delta \varphi
+ {\partial \zeta \ov \partial r}{\partial \varphi \ov \partial r} \r) +
{\partial \zeta \ov \partial t}{\partial \varphi \ov \partial t} - q_\pi
{\alpha(\varphi)a^4(\varphi) N^2 \ov 2}\bigl(\tilde{E} - 3\tilde{p}-
(\tilde{E}+\tilde{p})U^2\bigr) \label{e:boxfi}
\ee 
One more equation concerning the scalar field will be used; although it is 
redundant with (\ref{e:dmdr})--(\ref{e:boxfi}), from which it is deduced
, it will be useful for
numerical integration:
\be
{1\ov 2}{\partial \Xi \ov \partial t} = \l\{{N\ov A} \l[\psi \Delta \varphi
+ \eta {\partial \psi \ov \partial r}\r] + {2N\ov A} \psi \eta {\partial
\nu \ov \partial r} - \Xi {G_* A^2\ov r c^2}{\partial m\ov \partial t}
\r\} + \psi q_\pi\alpha(\varphi)a^4(\varphi) N\bigl(\tilde{E} - 3\tilde{p}-
(\tilde{E}+\tilde{p})U^2\bigr) \label{e:dxidt}
\ee

\subsection{Matter evolution equations}

In order to get the evolution of the variables $\tilde{E}$ and $U$, let us
consider the momentum-energy conservation equation (\ref{e:consen}), we get
\begin{eqnarray}
{\partial \tilde{E} \ov \partial t} + {1\ov r^2} {\partial \ov \partial r}
\bigl(r^2(\tilde{E}+\tilde{p})V\bigr) = -(\tilde{E}+\tilde{p})\l\{\alpha(
\varphi)N\l[(3+U^2)\psi+4U\eta\r]+rAN\l[(1+U^2)\psi\eta+U\Xi\r] \r\} 
\label{e:dedt} \\
{\partial U \ov \partial t} + V{\partial U \ov \partial r} = -{1\ov \tilde{E}
+\tilde{p}}\l(U{\partial \tilde{p}\ov \partial t}+{N\ov A}{\partial\tilde{p}
\ov \partial r}\r)-{AN\ov\Gamma}\l[{G_*m\ov r^2c^2}+q_\pi a^4(\varphi){r
\tilde{p}\ov 2}+{\alpha(\varphi)\ov A}\bigl(\eta+U\psi\bigr)+rU\eta\psi
+{r\ov 2}\Xi\r] \label{e:dudt}
\end{eqnarray}
Equation (\ref{e:dedt}) expresses the total energy conservation (matter plus
gravitational and scalar energy), eq.~(\ref{e:dudt}) being the tensor-scalar
analogous of the Euler equation. One notes, in the latter, the ${1\ov 
\tilde{E}
+\tilde{p}}\l(U{\partial \tilde{p}\ov \partial t}+{N\ov A}{\partial\tilde{p}
\ov \partial r}\r)$ term which may cause some trouble when numerically 
calculating it, since it is the quotient of two quantities vanishing at
the surface of the star. Thus, if one uses the log-enthalpy (\ref{e:defenth}),
this term may be replaced by ${1\ov\Gamma^2}\l(U{\partial H\ov \partial t}+
{N\ov A}{\partial H\ov \partial r}\r)$, which is well defined near the 
surface. 

Expressing the baryonic number conservation
\be
\tilde{\nabla}_\mu \tilde{n}_B \tilde{u}^\mu = 0
\ee
one obtains
\be
{\partial  \tilde{D} \ov \partial t} + a(\varphi){1\ov r^2}
{\partial\ov \partial r}
\l(r^2\tilde{D}V\r)+\alpha(\varphi)\tilde{D}N(3\psi + 4U\eta) = 0
\label{e:dddt}
\ee
Because log-enthalpy is used in order to avoid numerical
singularities at the surface, there has to be an evolution equation of that
quantity. Since $H = H(\tilde{n}_B)$, and $\tilde{n}_B = {\tilde{D}\ov A
\Gamma}$, one may write
\begin{eqnarray}
{\partial H\ov \partial t}+V{\partial H\ov \partial r}&=&
{\partial H\ov \partial \tilde{n}_B}\l({\partial \tilde{n}_B\ov \partial t}
+V{\partial \tilde{n}_B\ov \partial r}\r) \nonumber \\
&=&{\partial H\ov \partial \tilde{n}_B}{V\ov A\Gamma}\l(1-a(\varphi)\r)
{\partial  \tilde{D} \ov \partial r}-\tilde{n}_B{\partial H\ov \partial
\tilde{n}_B}\l[{1\ov r^2} {\partial\ov 
\partial r}(r^2V)+\alpha(\varphi)N(3\psi +4 U\eta) \right. \nonumber \\ 
& & \left. +{1\ov A}\l({\partial A\ov \partial t}+V{\partial A\ov 
\partial r}\r)
+ \Gamma^2U\l({\partial U \ov \partial t} + V{\partial U \ov \partial r}\r)\r]
\label{e:dhdt}
\end{eqnarray}
with terms in the right-hand side being replaced using eqs.~(\ref{e:dmdr}), 
(\ref{e:dmdt}) and (\ref{e:dudt}) by source terms involving $\tilde{E}, 
 \tilde{p} \text{ and }\Xi$. The results of numerical integration of
all these equations (\ref{e:dmdr})--(\ref{e:dhdt}) will be presented
in next section. 

\section{Numerical results}
\label{s:resu}

The numerical procedure, the code and its tests are described in
Appendix~\ref{sa:num}. In this section, only results are presented and
discussed.  An important choice is that of the coupling
function $a(\varphi)$. Following \cite{DEF96}, we chose a function
depending on two parameters, for all our study:
\be
a(\varphi) = e^{\alpha_0(\varphi-\varphi_0) + {\beta_0\ov 2}
(\varphi-\varphi_0)^2} \label{e:cfunc}
\ee
Figure~9 of \cite{DEF96} gives constraints on the ($\alpha_0, \beta_0$)
space of parameters, imposed by binary-pulsar
measurements. Section~\ref{s:expl} investigates this space of
parameters for scalar gravitational waves. Note that Brans-Dicke
theory is obtained for $\beta_0=0$.

\subsection{Static configurations}
\label{s:stat}

\begin{figure}
\centerline{ \epsfig{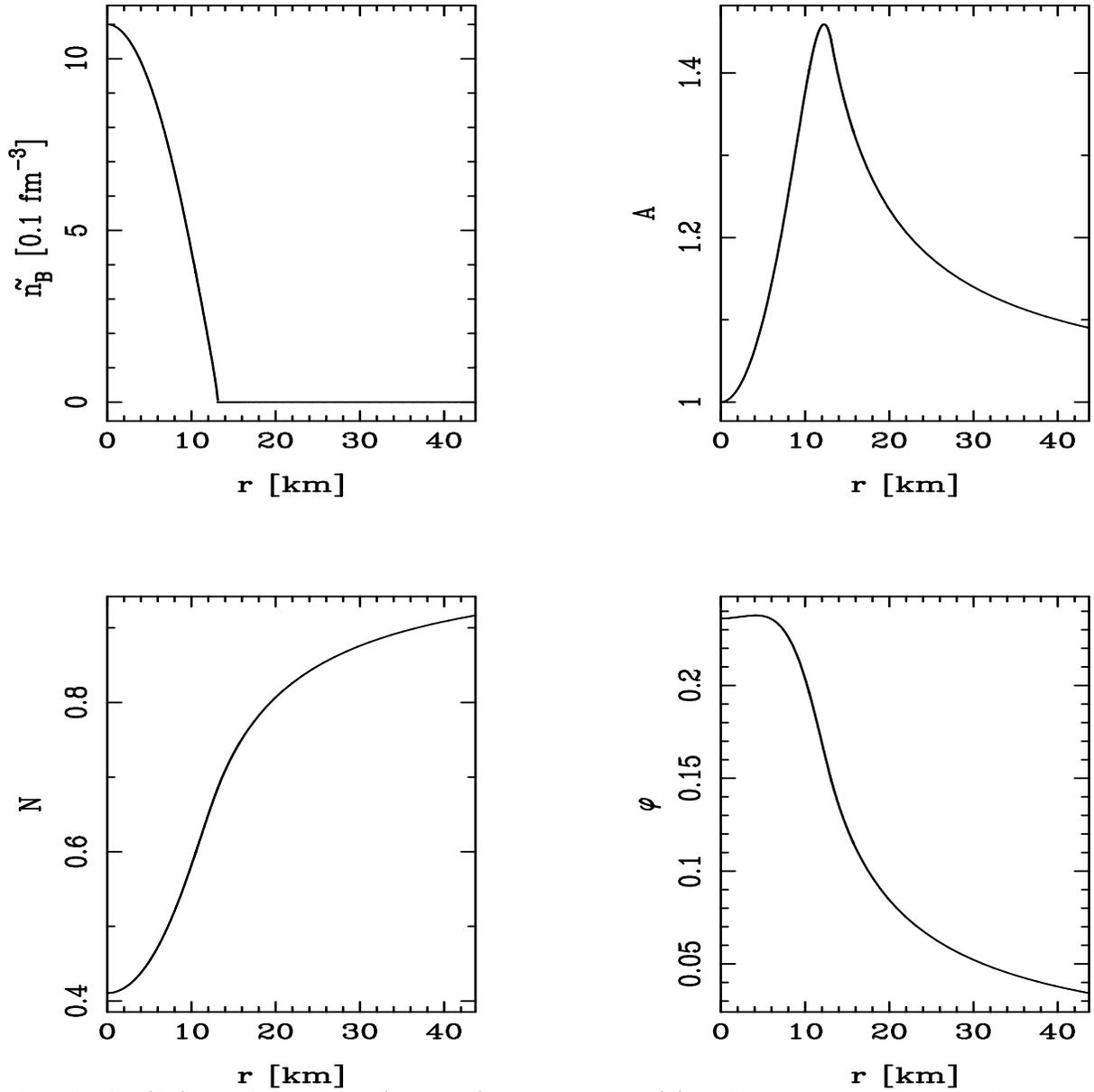} }
\caption[]{\label{f:srel}   
Density ($\tilde{n}_B$), metric potentials ($A$ and $N$) and scalar
field ($\varphi$) profiles for a neutron star of
 $2.4 M_\odot$, for EOS1 ($\gamma=2.34$ and $K=0.0195$ polytrope) and
with a coupling function $a(\varphi)=\exp(-3\varphi^2)$. The
asymptotic scalar field value is $\varphi_0=10^{-5}.$ Star's radius
$R_{star} = 13.1$ km.}
\end{figure}

Physical scenarios to form a black hole involve either an accreting
neutron star or a post-supernova remnant (when a part of the ejected
envelop falls back onto the new-born neutron star).
In both cases, the mass of the neutron
star must reach its maximal value above which the star becomes
unstable. It is then interesting to get unstable equilibrium configurations
of neutron stars, endowed with a scalar field, close to the maximal
mass. They are used as initial configurations for the collapse. Thus,
setting all $\partial / \partial t$ terms to zero, as well as $V$ and $U$ in
eqs.~(\ref{e:dmdr})--(\ref{e:dedt}), one gets the scalar equivalent of
the Tolman-Oppenheimer-Volkoff system.  The system obtained is the
same as equations (7) in \cite{DEF93}, since the same gauge is
used. 
Considering a polytropic equation of state
\begin{eqnarray}
\tilde{e}(\tilde{n}_B) = \tilde{n}_B \tilde{m}_B+K{\tilde{n}_0\tilde{m}_B
\ov \gamma -1}\l({\tilde{n_B} \ov \tilde{n}_0}\r)^\gamma \nonumber \\
\tilde{p} = K \tilde{n}_0\tilde{m}_B\l({\tilde{n_B} \ov \tilde{n}_0}
\r)^\gamma
\end{eqnarray}
with $\tilde{m}_B=1.66 \times 10^{-27} \text{ kg}$ and 
 $\tilde{n}_0 = 0.1 \text{ fm}^{-3}$, 
one can integrate the scalar TOV system, starting at the
center with a given value for $\tilde{n}_B(r=0)$, up to the surface
at which $\tilde{n}_B(r=R_{star})=0$. In this study, two types of
polytrope will be used:
\begin{enumerate}
\item $\gamma=2.34$ and $K=0.0195$, which has already been used by
\cite{DEF93} to fit EOSII in \cite{DAI85}; called EOS1 in this paper.
\item $\gamma=2$ and $K=0.1$, as used in \cite{SBG94}; called EOS2 in
this paper.
\end{enumerate}
Then for each static configuration, the total ADM mass of $g^*_{\mu\nu}$,
which will be called {\em gravitational mass}\ and the total scalar charge
 $\omega$ such that, for $r \rightarrow \infty$, 
$\varphi(r) = \varphi_0 +G_*\omega /r + O(1/r^2)$,
 can be determined through eqs.(8) of \cite{DEF93}.
These two quantities are useful to match the obtained interior
solution to the exterior one (spherically symmetric solution in vacuum), which
is known analytically in an other gauge (described in \cite{CEF90}) and thus
one can obtain a static solution everywhere. The resulting fields are
shown in Fig.~\ref{f:srel} for $\beta_0=-6$, with large value of the
scalar field inside the star, even for a very small asymptotic field
$\varphi_0$ (spontaneous scalarization). These solutions are then used
as initial values for the dynamical evolution.  They have been
computed, with increasing central densities $\tilde{n}_B(r=0)$, in order to
get an ``unstable'' configuration (for which gravitational mass is a decreasing
function of the density). This property is not evident in
tensor-scalar theory, but the dynamical code being
sensitive enough to trigger the instability only by round-off errors,
it has been checked numerically.
The hydrostatic equilibrium is obtained,
thanks to pseudo-spectral techniques (see Appendix~\ref{sa:cheb} and
\cite{BOM86}), up to very high accuracy 
($10^{-10}$ relative error on the hydrostatic equilibrium), which
enables the dynamical code to be
sensitive to instability (see \cite{GOU91}).

\subsection{Scalar gravitational waves}
\label{s:scagw}

\begin{figure}
\centerline{ \epsfig{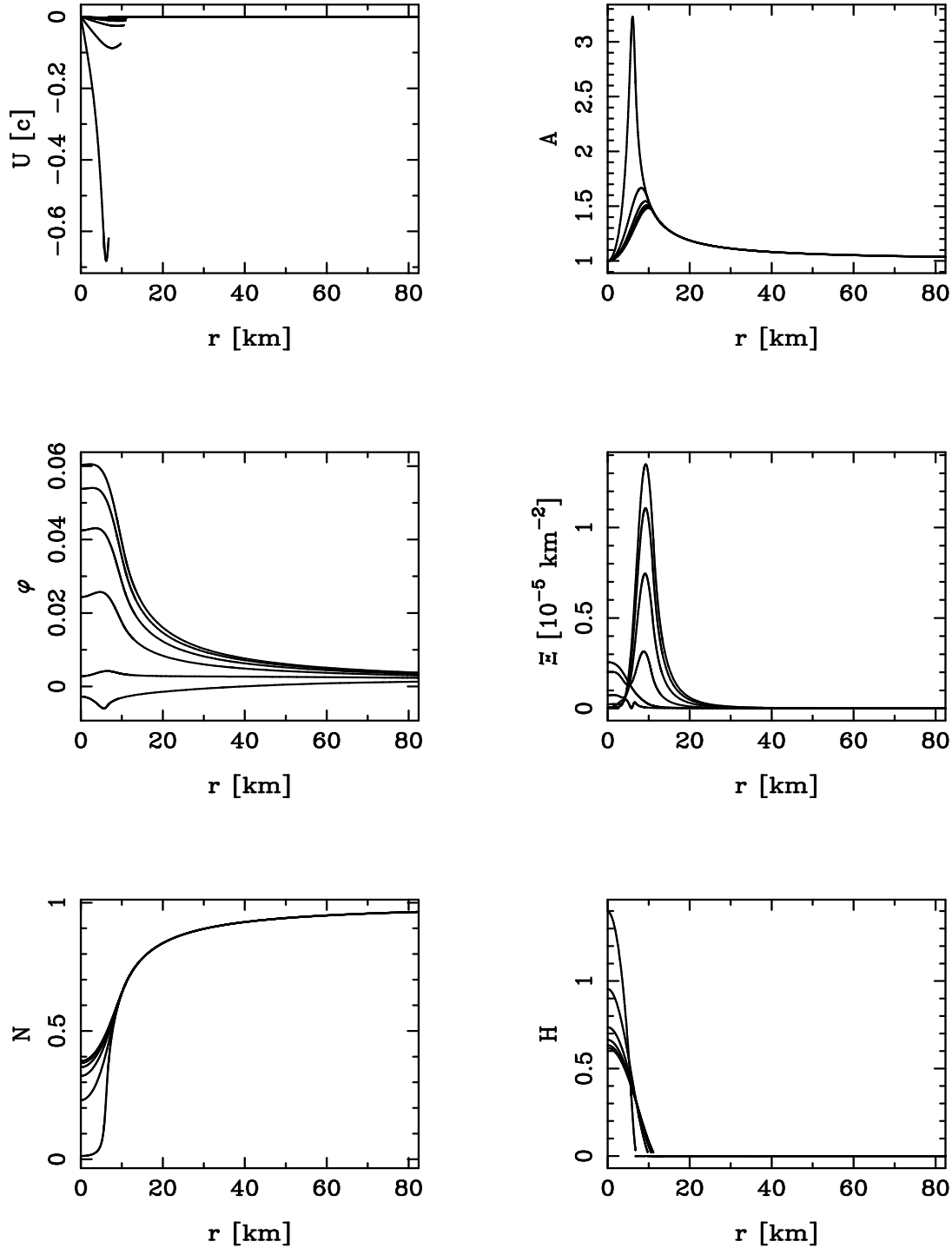} }
\caption[]{\label{f:dyncol}
Profiles of various quantities at different values of $t$ between
 $0$ and $4.64$ ms, for collapse A. The fluid velocity $U(r,t)$, measured
by the hypersurface observer, is expressed in units of $c$ 
and its evolution is downward, the extremity of each curve giving the
position of the star's surface at the corresponding instant. The
evolution for $A(r,t)$ (metric potential), $\Xi(r,t)$ (scalar ``energy'') and
 $H(r,t)$ (log-enthalpy) is upward, for $\varphi(r,t)$ (scalar field)
and $N(r,t)$ (lapse) is downward. }
\end{figure}

\begin{figure}
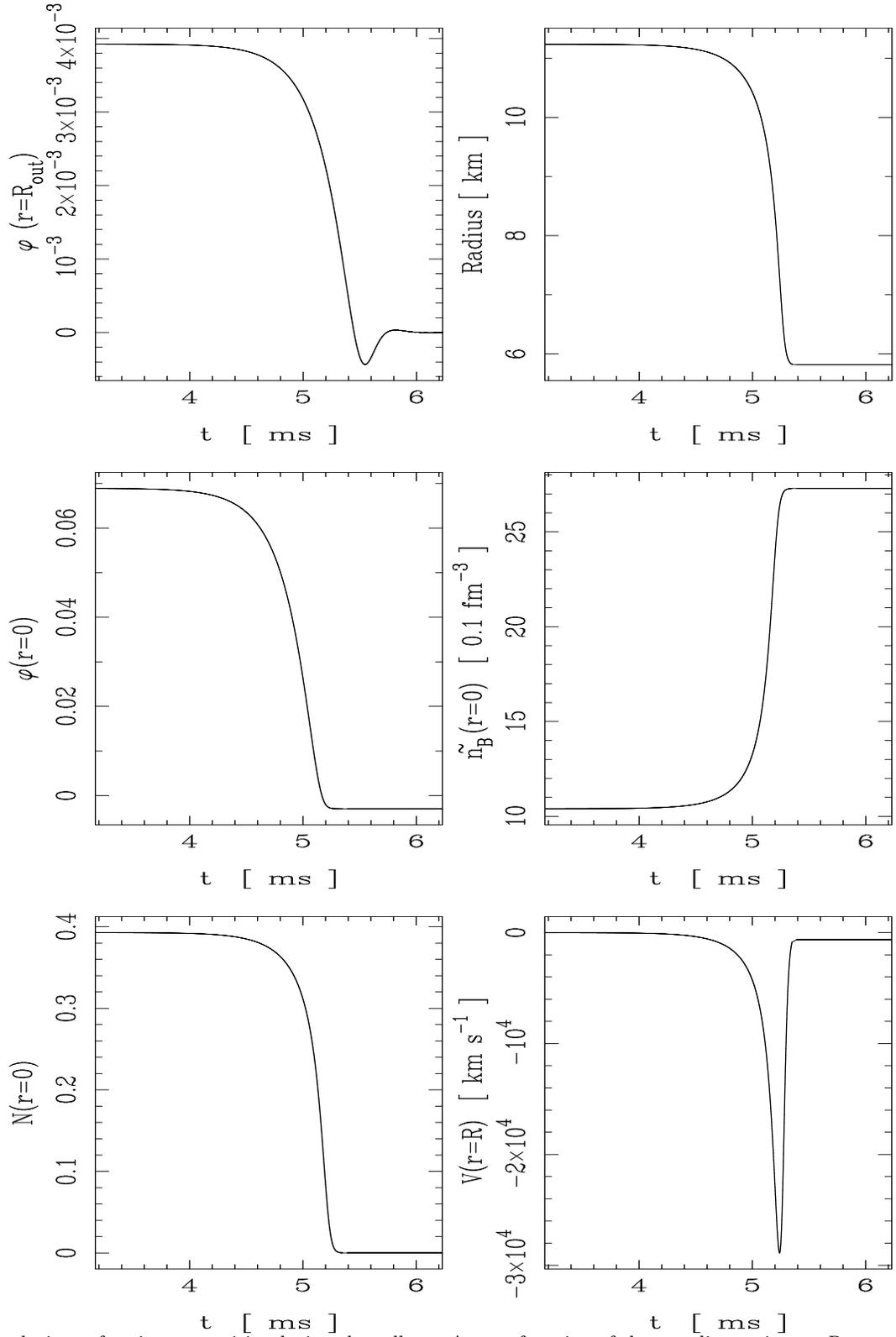

\centerline{\epsfig{figure=f11b.ps,height=7cm,width=7cm,angle=0}
\epsfig{figure=f14b.ps,height=7cm,width=7cm,angle=0}}
\centerline{\epsfig{figure=f15b.ps,height=7cm,width=7cm,angle=0}
\epsfig{figure=f16b.ps,height=7cm,width=7cm,angle=0}}
\centerline{\epsfig{figure=f17b.ps,height=7cm,width=7cm,angle=0}
\epsfig{figure=f18b.ps,height=7cm,width=7cm,angle=0}}
\caption[]{\label{f:gps306}
Evolutions of various quantities during the collapse A, as a function
of the coordinate-time $t$. $R_{out}=300\text{ km}$ is
the radius of the outer edge of the grid. $\varphi(r,t)$ is the scalar
field and the radius is the coordinate value for which $\tilde{n}_B$, the
baryon density is zero. $N(r=0)$ is the lapse at star's center and
$V(r=R)$ is the star's surface velocity.}
\end{figure}

Hereafter, four collapse calculations  will be presented, called A, B,
C and D. The 
parameters of the static configurations, which were used as initial
conditions for the collapses, are described in table~\ref{t:coll}. Note
that collapses A and B use a $\gamma=2.34$ polytrope, whereas C and D
use a $\gamma=2$ one.
 First, only the case A will be considered. As far as the
hydrodynamic part is concerned, the collapse is very
similar to that in general relativity, described in
\cite{GOU91}. It can be seen from Fig.~\ref{f:dyncol}, that
 $A(t,R_{star}(t))\to\infty$ due to the pathological behavior of the
radial gauge when $R_{star}$ is approaching the Schwarzschild
radius. An apparent horizon is expected to develop, but minimal
2-surfaces cannot be described by the radial gauge. On 
Fig.~\ref{f:dyncol} are plotted several quantities during the
collapse (until $N(r=0)$ becomes too small). Thus, although RGPS
coordinates are not well adapted for the description of a black hole,
they were used to describe the collapse toward it, as in
\cite{GOU91}. Moreover, from Figs.~\ref{f:dyncol} and \ref{f:gps306}
one sees that the star has almost entered its Schwarzschild radius
(${R_{star}\ov R_{Schwarzschild}}=1.001$ at the end of the collapse);
so that no significant later evolution could be achieved inside the
star. Actually, one notices that the lapse goes to zero within the
Schwarzschild radius of the star. Since all evolution equations are
written $\partial / \partial t = N \times (\text{source term})$ and
the coordinate velocity $V = \l(N/A\r)\, U\to 0$, all hydrodynamic and
scalar-field quantities are ``frozen'' inside the star. Therefore,
their evolution can be numerically stopped, in order to avoid the
singularity of $A(r=R_{star})$. However, all field quantities continue
to evolve {\it outside} the star, as long as one wants in terms of
coordinate time (which is the time of an observer at spatial
infinity).

\begin{figure}
\centerline{\epsfig{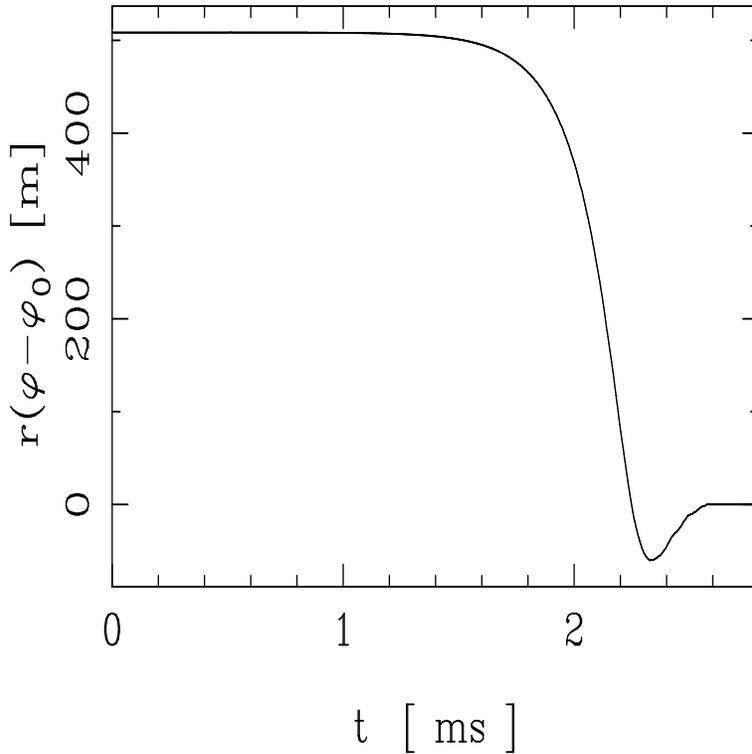}}
\caption[]{\label{f:wvform}
Waveform of the emitted signal during collapse B. The plotted quantity
is the function $F(t)$ (see sec.~\ref{s:scagw}), measured at
 $r=300\text{ km}$ and expressed in meters. $\varphi(r,t)$ is the
scalar field and $\varphi_0$ its asymptotic value. To get the gravitational
wave amplitude $h$, one must use eq.~(\ref{e:ampli}).}
\end{figure}

The results of this evolution are shown in Fig.~\ref{f:gps306}, for
collapse A. The fate of the scalar
field is particularly interesting: for $r\gg R_{star}$ the field
relaxes toward the asymptotic constant value set by cosmological
evolution; the scalar energy of the star is radiated away as scalar
gravitational wave. The scalar field ($\varphi(r=R_{out})$) is considered
to be sufficiently far away from the star (i.e. in the wave zone) to
give the monopolar gravitational wave signal. Using the ``frozen
star'' to evolve the fields outside the star, makes the integration
time long enough to get all the information from the collapse at 
$r\gg R_{star}$ (where all gauges become equivalent). The main difference
from general relativity is the scalar monopolar radiation, which
carries away energy and can interact with a detector. Looking far from
the source (at a distance $r\gg R_{star}$), one can write the metric (see
\cite{DEF92})
\be
\tilde{g}_{\mu\nu}(r,t) = a^2(\varphi_0) \l[ f_{\mu\nu} + {1\ov r} \l(
h_{\mu\nu} + 2\alpha_0 F f_{\mu\nu} \r) + O(r^{-2})\r] 
\ee
 where $f_{\mu\nu}$ is the flat metric, $h_{\mu\nu}(t-{r\ov c})$ and
$F(t-{r\ov c})$ are
respectively the quadrupolar and monopolar components of the
wave. Since this work is done in spherical symmetry, only the
monopolar mode shall be considered. The function $F$ is the same as
that of eq.~(\ref{e:ccfi}), and is plotted for collapse B in
Fig.~\ref{f:wvform}. If one wants to compare the amplitude to those of 
general relativistic gravitational waves ($h^{TT}_{ij}$), at a
distance $d$, then the right quantity is (see \cite{DEF92},
\cite{HCN97} or \cite{WK97}):
\be
h(t) = {2 \ov d} a^2(\varphi_0)\alpha_0 F(t)
\label{e:ampli}
\ee

\begin{figure}
\centerline{\epsfig{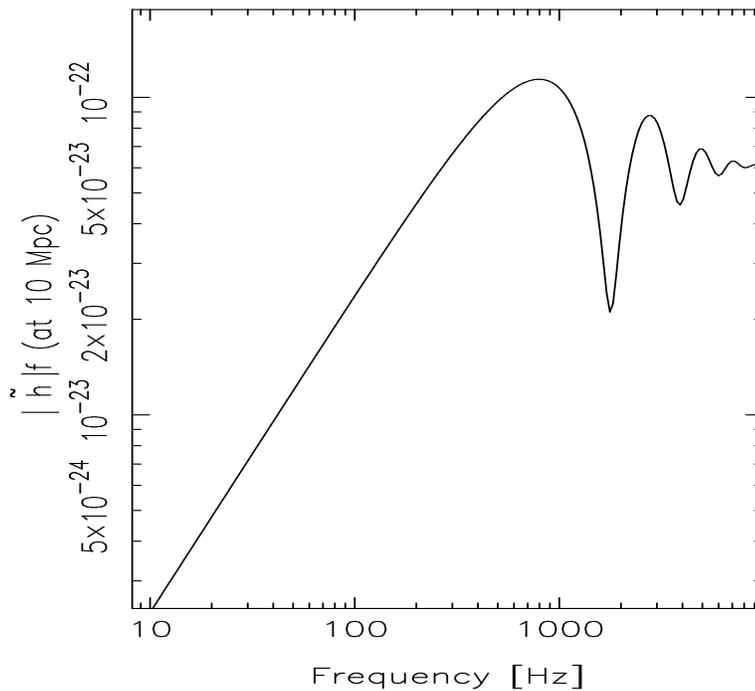}}
\caption[]{\label{f:fou356}
Fourier power spectrum of the scalar gravitational wave emitted during
collapse B. $\tilde{h}$ is the Fourier transform of the signal,
measured at $10$ Mpc and $f$, the frequency.}
\end{figure}

The Fourier spectrum of the signal is 
\be
\tilde{h}(f) = \int_0^{t_{max}} h(t) e^{2\pi i ft} dt
\ee
and the power spectrum ($f|\tilde{h}|$), which is plotted in
Fig.~\ref{f:fou356} for the collapse B, is then useful to determine
the characteristic frequency $f_c$ of the signal. Actually, the
quantity to be compared with detectors' sensitivity is 
\be
\tilde{h}_c = \l | \tilde{h}(f_c)\r|\sqrt{f_c} \label{e:hc}
\ee
which is expressed in $\text{Hz}^{-1/2}$. From Fig.~\ref{f:fou356},
one sees that a ``collapse B'' could be detected by VIRGO, provided it 
occurs within 300 kpc.
Finally, the radiated Einstein-frame Bondi energy writes
\be
E_{scal} = {c^3\ov G_*} \int_0^{+\infty} \l({dF\ov dt}\r)^2 dt \label{e:eray}
\ee
which is the total energy radiated in gravitational waves.

\subsection{Comparison with previous works}
\label{s:compar}

Because all previous works have studied an Oppenheimer-Snyder
collapse, for comparison some runs were done putting $\tilde{p}$ and
$H$ to zero.

\begin{figure}
\centerline{\epsfig{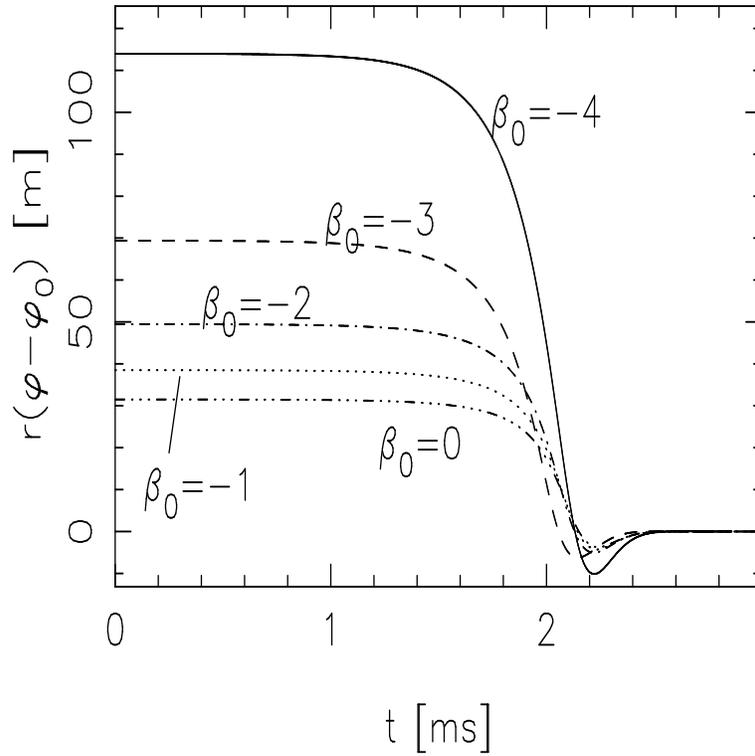}}
\caption[]{\label{f:db0m4}
Same as Fig.~\ref{f:wvform} but for different
collapses. $\alpha_0=3.16\times 10^{-2}$ for all of them, but
 $\beta_0$ varies from $-4$ (upper curve, ---) to $\beta_0=0$
(lower curve, $-\cdots-$).}
\end{figure}

The first work by Shibata et al. \cite{SNN94} considered the dust
collapse in the Brans-Dicke theory ($\beta_0 = 0$). The authors used two
types of initial conditions, called $(A)$ and $(B)$. The case $(A)$
starts the collapse with $\varphi = \varphi_0$, whereas $(B)$ starts
it with a quasi-static solution for $\varphi$. Making the same dust
collapses, the same waveforms and amplitudes were obtained (their
 $\Phi$ is related to $\varphi$ by $\Phi = 2\log\l(A(\varphi)\r) =
 2\alpha_0(\varphi-\varphi_0)$). Taking the equation of state into
account, with equilibrium initial configurations (which are the most
realistic possible), gives the waveform of Fig.~\ref{f:db0m4}
($\beta_0=0$). The form and amplitude are very close to the $(B)$ type
collapses of Shibata et al.(see their Fig.~3), the power spectra being
close too (Fig.~\ref{f:fou356}, with lower amplitude, and Fig.~6 of
\cite{SNN94}). 

Another work in Brans-Dicke theory, with dust matter, was done by
Scheel et al. \cite{SST95}, dealing more with the fate of the final
object. However, comparing $2\alpha_0 dF/dt$ of this work, with
$f'$ of their paper gives the same result. Whereas the authors can
conclude on the final state of the collapse, this work cannot describe
it. All that can be said is that the scalar charge (or mass) is all
radiated away, so that the scalar field relaxes toward its
cosmological value, and that an apparent horizon is {\it expected} to
appear (Cf. sec.~\ref{s:scagw}). These results still hold in more
general tensor-scalar theory. Finally, it can be pointed out that,
contrary to their work, here the tensor mass in the {\it Einstein}
frame is considered, not in the Jordan-Fierz one.

\begin{figure}
\centerline{\epsfig{figure=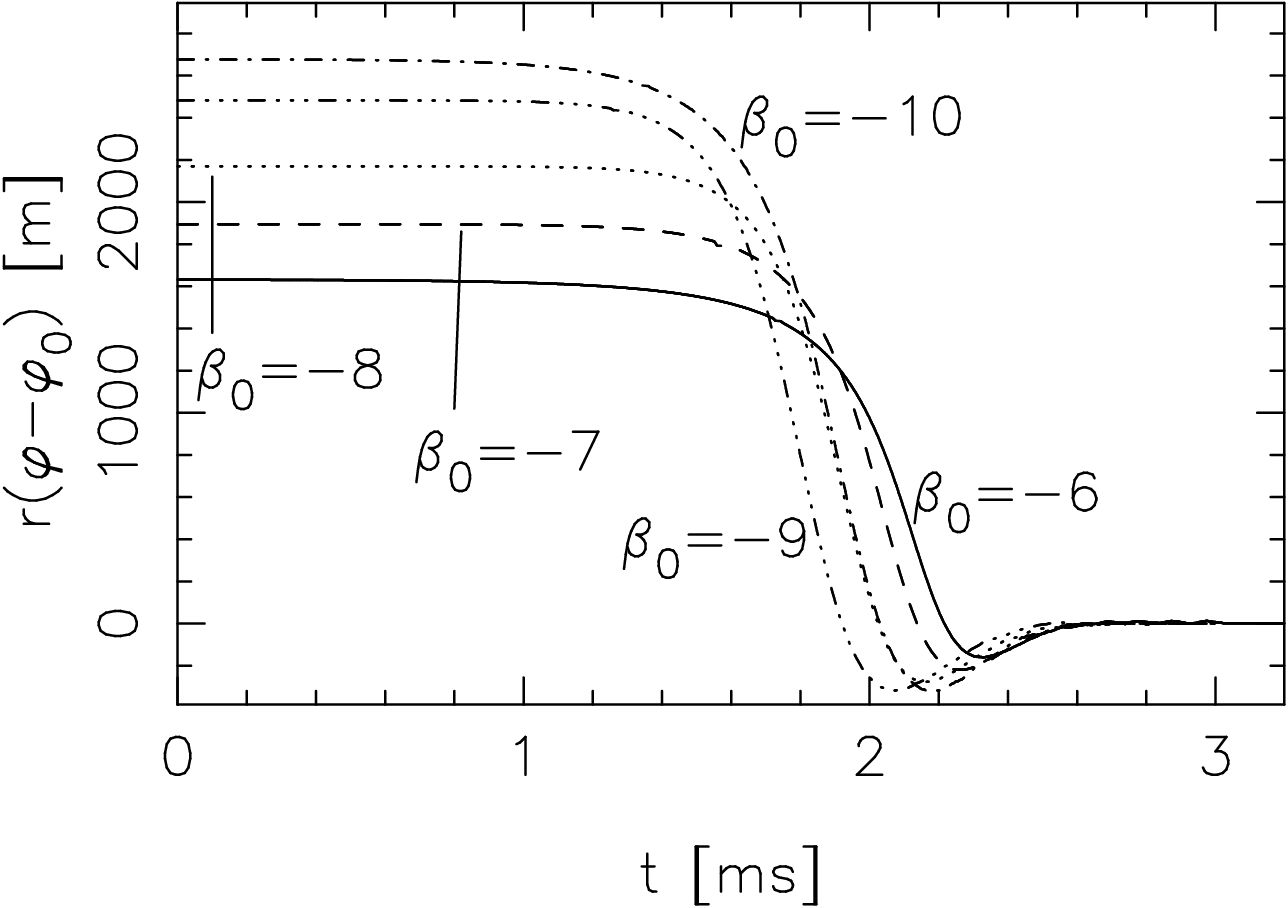,height=10cm,width=10cm,angle=270}}
\caption[]{\label{f:dbm6m10}
Same as Fig.~\ref{f:db0m4} but
 $\beta_0$ varies from $-10$ (upper curve, $-\cdot-$) to $\beta_0=-6$
(lower curve, ---).}
\end{figure}

The last study of spherically symmetric collapse in tensor-scalar
theory was done by Harada et al. \cite{HCN97}. They used the
Oppenheimer-Snyder metric (in general relativity) as a background
spacetime and did some Taylor expansion of tensors and equations in
terms of scalar field coupling function parameters. Therefore, they
used unrealistic initial conditions ($\varphi = \varphi_0$) and could
not study the cases of spontaneous scalarization. Waveforms
resulting from our calculations for different $\beta_0$, from $-10$ to
 $50$ are shown in Figs.~\ref{f:db0m4} to \ref{f:db4050}, with
unstable equilibrium initial configurations and full
hydrodynamics. $\alpha_0=3.16\times 10^{-2}$ as in \cite{HCN97} and
the mass is the maximal one (Cf.\ref{s:stat}), the ratio
 $R_{star}/M\simeq 4$. The results are different, although
showing the same tendancy as in \cite{HCN97}, for many
 $\beta_0$, due to the fact that Harada et al. took unrealistic initial
conditions. Thus in their simulations, when the collapse begins, the
scalar field  on the one hand evolves to reach its quasi-equilibrium
configuration, on the other hand, it feels the effects of the
collapse. Their signal is then a superposition of these two effects:
raise of the scalar field up to its equilibrium value, with one or
several oscillations depending on $\beta_0$, then the fall of this field,
due to the collapse of the matter. The signals for $\beta_0>20$ have
more important oscillations than other ones; this may be explained as
follows. Inside the star, there are unstable modes which develop for
these $\beta_0$ and when the ratio $R/M$ becomes small, as it has
been shown by Harada in for static configurations (see
Fig.~5 of \cite{HAR97}). This is due to the fact that, near the star's
center, $\tilde{E}-3\tilde{p} <0$ allowing for spontaneous
scalarization to develop\footnote{see the simplified model of
\cite{DEF93}}. Here, one can see an effect of the pressure on the
signal. These non-perturbative modes develop on a time-scale $\tau$
$$
\tau \sim {\tau_{\text{ff}} \ov \sqrt{|\beta_0|}}
$$
(see eq.~(5.3) of \cite{HAR97}) with $\tau_{\text{ff}}$ being the
free-fall time of the star. Thus, for $\beta_0>20$ the modes develop
during the collapse, when $R/M$ becomes small enough. This rise is
fast enough to be (at least partly) seen before the star enters its
Schwarzschild radius (see Figs.~\ref{f:db1030} and \ref{f:db4050}). 

\subsection{Exploring the parameter space}
\label{s:expl}

Thanks to the fact that it solves complete equations, the code
presented in this paper is able to explore a larger part of the parameter
space and give more ``physical'' results. In sec.~\ref{s:compar}
$\beta_0$ has been varied from -10 to 50 and $\alpha_0$ was fixed. The
effect of spontaneous scalarization, for $\beta_0\simeq -5$ and
lower, was observed changing the amplitude, but not the shape of the
wave. However, the two regimes (depending on the value of $\beta_0$)
were studied separately, when varying $\alpha_0$. This latter has been 
taken between $0$ and $3\times 10 ^{-2}$, either in spontaneous
scalarization ($\beta_0= -6$), or without it ($\beta_0 = -4$). The
results are shown in Fig.~\ref{f:dfidal} for $\tilde{h}_c$ and in
Fig.~\ref{f:erydal} for $E_{scal}$. One notices that for $\beta_0=-6$
$\tilde{h}_c\propto \alpha_0$ and $E_{scal} \propto \text{constant}$,
whereas both $\propto \alpha_0^2$ for $\beta_0=-4$. The scalar field
amplitude does (almost) not depend on $\alpha_0$ in spontaneous
scalarization, whereas it is directly proportional to it otherwise,
like in Brans-Dicke theory.

The equation of
state has also been changed. Collapse C and D have  
been performed with EOS2 (see sect.~\ref{s:stat}). Results are shown in
Fig.~\ref{f:fou256} for collapse D and are similar to those of
EOS1 (collapses B), the scalar gravitational wave signal having
the same shape, shifted since $f_c$ is higher ($1$ kHz for D, versus
$800$ Hz for B). This difference as well as that in amplitude can easily
be explained by the change of mass of unstable-equilibrium
configuration. Thus, varying the parameters of
the polytropic equation of state is not of much interest. Better would
be the use of a
realistic equation of state (as in \cite{GOU91}) but, for this work, a
polytrope gives already good results.

Finally, a few more runs were performed, with EOS1, varying $\beta_0$
from $50$ to $-10$ and, for each $\beta_0$, the maximal value of
$\alpha_0$ allowed by solar-system experiments (see e.g. Fig.~9 of
\cite{DEF96}) was taken. More precisely, the lowest value of
\be
\alpha_0^2<10^{-3}\text{ and } \alpha_0^2<{1.2\ov |\beta_0|}\times
10^{-3} \label{e:alfamax}
\ee
was taken. Results are shown in Figs~\ref{f:eray1} and \ref{f:eray2}
for $E_{scal}$, and it can be seen that effects of spontaneous
scalarization appear for $\beta_0<-4.4$ and that, for
$\beta_0>20$, one can see effects of instabilities of the scalar field 
(see section~\ref{s:compar}). As far as the gravitational wave signal
is concerned, the quantity plotted in Figs.~\ref{f:deltfi1} and
\ref{f:deltfi2} is $\tilde{h}_c$, at $10$ Mpc from the source. The
characteristic frequency range is $700\text{ Hz}\leq f_c\leq 900$ Hz 
except for $\beta_0 \geq 20$, where $f_c \sim 2000$ Hz. The maximal signal
 $\tilde{h}_c \simeq 5\times 10^{-24}$ at $10$ Mpc which is, at least,
one order of magnitude lower than the best expected sensitivity of
interferometric detectors currently under construction (see
e.g. \cite{VIR2}). However, one may compare the signal amplitude and
energy to those of similar collapses in general relativity (2 and 3\,D
stellar core collapse, see \cite{BOM93} and \cite{ZMU97}) and see that
both are quite higher. Thus, if spontaneous scalarization effects are
likely to occur, their gravitational signal should be more easily
detected than the quadrupolar one from a collapsing source (aborted
supernova or neutron star reaching its maximal mass by accretion).

\section{Conclusions}
\label{s:conc}

This work has been done with very few approximations ($A\times N$ set
to a constant at the outer edge of the grid, see eq.~(\ref{e:AN}) and
Appendix~\ref{sa:dyn}, evolution ``frozen'' when the
lapse becomes too small); all the tensor-scalar equations, including
hydrodynamics, were solved with  high accuracy by means of spectral
methods. Although the gauge choice does not allow any study of the
state of the final object, the scalar gravitational wave signals can be
obtained and the results compare well with previous (simplified)
works. From that comparison it can be said that, whereas taking
pressure into account, in most cases, does not have significant
effects on the signal (which was to be demonstrated), it is the only
way to get reliable initial conditions and to trigger the collapse in
a ``natural'' way. Doing so helps to get clean waveforms. On the other
hand, since the complete set of tensor-scalar equations was solved, it was
possible to study the effects of coupling function parameters. Mainly, 
one sees that the outgoing monopolar gravitational wave is very
dependent on the coupling function, especially the $\beta_0$
parameter. This is interesting because the $\alpha_0$ parameter can be
constrained by solar-system experiments, since it represents only
a linear deviation from general relativity, whereas $\beta_0$ cannot be
really probed by that mean. Even if the signal from extra-galactic
sources is not strong enough to be detected, it is higher than the
quadrupolar one and involves more energy. It means that, if a
quadrupolar wave signal is detected by VIRGO or LIGO, with no
monopolar component, then the constrain on tensor-scalar theory will
be quite strong. A
future project of study is the supernova collapse and bounce in this
framework since, in that case, one has electromagnetic and neutrino
signals which make possible the use of even negative results of detection.
 
\acknowledgments

I thank Thibault Damour for suggesting this work, for fruitful
discussions and reading of the manuscript. I am very grateful to Eric
Gourgoulhon for help, discussions and careful reading of the
manuscript. The numerical calculations have been performed on Silicon
Graphics workstations purchased thanks to the support of the SPM
department of the CNRS and the Institut National des Sciences de l'Univers.

\appendix

\section{Numerical procedure}
\label{sa:num}

Hereafter, some details of the numerical techniques (spectral methods) are
described. More complete explanations can be found in \cite{BOM86},
\cite{BOM90} and \cite{BGM97}.

\subsection{Chebyshev decomposition}
\label{sa:cheb}

The numerical problem is to solve a set of partial differential
equations. For this purpose, each field (or function) $f_{t_0}(r)$ is
represented, at a given time $t_0$, as a truncated series of Chebyshev
polynomials (or as a column vector of the coefficients of this
series). Usual number of coefficients (or points) is between 17 and
65. Then all spatial operators like
$$
f\rightarrow {\partial f\ov \partial r},\ \int f,\ {1\ov r}f,\ rf,\
\Delta f
$$
reduce to matrix multiplications of the set of $f$'s
coefficients. Constraint equations (\ref{e:dmdr}) and (\ref{e:dnudr}) are
thus easily integrated. Other operations, like multiplication of two
functions, must still be done in the physical space (at grid's
points).

Evolution equations are written in the form
\be
\l.{\partial f\ov \partial t}\r|_{t=t_{J+1/2}}={f_{t_{J+1}}-f_{t_J} \ov
t_{J+1} -t_J}={\cal S}(f)_{t_{J+1/2}}
\label{e:intg}
\ee
with $\cal S$ being a spatial operator on $f$ and $t_J$ being the
$J$-th instant of integration. There is need to evaluate $\cal S$ at time
 $t_{J+1/2}$, which can be done either explicitly (extrapolated from known
quantities at $t_J$ and $t_{J-1}$) or implicitly (interpolated from
unknown quantity at $t_{J+1}$). Explicit integration suffers from the
severe Courant-Friedrich-Levy constraint on the time step. However, in
the case of an advection equation (like (\ref{e:dddt})), it can be used
almost with an arbitrary time step provided that the advection velocity is
zero at the edges of the grid; hence the {\it comoving} grid with the
fluid (see Appendix~\ref{sa:dyn}). The implicit way requires to write
\be
\l(1-{(t_{J+1}-t_J)\ov 2}\cal{S}\r).f_{t_{J+1}}=f_{t_J} +
{(t_{J+1}-t_J)\ov 2}{\cal S}(f_{t_J})
\label{e:intimp}
\ee
where the function $f$ is represented by its coefficients and with
 $(1-{(t_{J+1}-t_J)\ov 2}\cal{S})$ being a matrix, which is inverted to
get the solution $f_{t_{J+1}}$. In the case where one has to 
impose boundary conditions, the right-hand side of eq.~(\ref{e:intimp}) is
replaced by a vector of coefficients containing zeros except on the
last or last but one column. One thus gets a ``free'' solution which
can be combined with the first one in order to satisfy boundary
conditions.

\subsection{Wave equation}
\label{sa:wave}

Unfortunately, the $\cal{S}$ operator present in eq.~(\ref{e:intg}) is not
always linear (i.e. not represented by a matrix in coefficient
space). That is the case for the wave equation (\ref{e:boxfi}), with a 
spatial operator of the form $\varphi \rightarrow e^{2\zeta}\Delta
\varphi$. This equation is then decomposed as follows:
\begin{eqnarray*}
{\partial ^2 \varphi \ov \partial t ^2}& = &\l(\lambda_2^{(t)}r^2 +
\lambda_1^{(t)}r + \lambda_0^{(t)}\r)\Delta \varphi + \sigma_\varphi \\
\text{with } &&\\
\sigma_\varphi & = &\l(e^{2\zeta}-\lambda_2^{(t)}r^2 +
\lambda_1^{(t)}r + \lambda_0^{(t)}\r)\Delta \varphi +e^{2\zeta}
{\partial \zeta \ov \partial r}{\partial \varphi \ov \partial r} +
{\partial \zeta \ov \partial t}{\partial \varphi \ov \partial t} +
\Bigl(\text{Matter source terms}\Bigr)
\end{eqnarray*}
and $\lambda_2^{(t)}r^2 +\lambda_1^{(t)}r + \lambda_0^{(t)}$ being an
approximation of $e^{2\zeta(r,t)}$, allowing to write the most
important part of the spatial operator in a ``linear'' form. Finally, one
writes $\l.\partial^2 \varphi /\partial t^2 \r|_{t=t_J}$ with
second-order approximation and $\Delta \varphi_{t_J}=(\Delta
\varphi_{t_{J+1}} + \Delta\varphi_{t_{J-1}})/2$ to make the
integration scheme implicit. The
boundary condition imposed on the outer edge of the grid (far away from
the star) is that of an {\em outgoing wave}, meaning that the wave can be
written as
\be
\varphi(t,r) = \varphi_0 + {1\ov r}F(t-{r\ov c}) \label{e:ccfi}
\ee
which is an exact condition for 1-D waves. Differentiating, one gets
\be
\l .
{1\ov c}{\partial \varphi \ov \partial t} + {\partial \varphi \ov
\partial r} + {(\varphi-\varphi_0) \ov r} \, \r |_{\text{outer edge}} = 0
\ee
Actually, this is not the right boundary condition for a wave equation
in curved space-time, however, since the boundary condition is imposed
far away from the star (i.e. on a nearly flat space-time), a good 
approximation is obtained by taking as a boundary condition
\be
\l.
e^{-\zeta}{\partial \varphi \ov \partial t} + {\partial \varphi \ov
\partial r} + {\varphi \ov r} \, \r |_{\text{outer edge}} = 0 
\label{e:ccont}
\ee

\subsection{Dynamical evolution} 
\label{sa:dyn}

The integration procedure is quite similar to that of Gourgoulhon in
\cite{GOU91}. All matter and field quantities are supposed to be known
at some initial instant $t_0$, and one wants then to get them at $t_0
+\delta t$. First, one can compute the scalar field variables $\varphi$
and $\Xi$ at that
time thanks to wave equation (\ref{e:boxfi}) and
eq.~(\ref{e:dxidt}). Similarly, one gets the fluid quantities
 $\tilde{E}$, $U$ (and thus $\Gamma$), $H$ and 
$\tilde{D}$, with their
evolution equations (\ref{e:dedt}), (\ref{e:dudt}), (\ref{e:dhdt}) and
(\ref{e:dddt}). Then, one
can deduce the metric coefficient $A(r,t_0+\delta t)$ through 
eq.~(\ref{e:dmdr}) and (\ref{e:defa}) (the integration constant is
obtained by the condition $\forall t,\ A(r=0,t)=1$); and determine the
fluid proper baryonic 
 density $\tilde{n}_B$, by inverting relation (\ref{e:defd}). The equation
of state then gives the pressure $\tilde{p}(\tilde{n}_B)$.

Finally, one uses eq.~(\ref{e:dnudr}) to obtain $\nu(r,t_0+\delta t)$, which
is determined up to an additive constant. Since there is no Birkhoff Theorem
in scalar-tensor gravity, this constant cannot be determined by matching the
interior solution to the exterior (static) one, used before. Even the exterior
space-time is dynamic. Using a large grid (which is going far away from the
star, typically $\sim 30\times R_{star}$) enables to be in weak-field
regime at the 
outer edge of the grid and so, to write with a good approximation
\be
\forall t, A(R_{out},t)\times N(R_{out},t) = K_{AN}
\label{e:AN}
\ee
$K_{AN}$ being a constant determined for the static configuration
($K_{AN}=1$ for General Relativity). One thus gets the integration
constant for $\nu(r,t_0+\delta t)$. Observing
 $A(R_{out})$ during the collapse, one sees that 
 $\Delta A(R_{out}) \leq  10^{-5}$. This is quite lower than the
overall committed error, see Appendix~\ref{sa:test}.
Once $N(r,t)$ is obtained, the velocity $V$ is deduced and all 
quantities are known at the instant $t_0+\delta t$. 

 Time integration is performed
by means of a second-order semi-implicit scheme (see previous sections),
boundary conditions being imposed on the $H,U$ system
\be
\l.{1\ov r^2}{\partial\ov \partial r} r^2 V\r|_{r=R_{star}} = 0
\ee
and on the wave equation (\ref{e:ccont}). Contrary to \cite{GOU91},
the code is working with several domains. Each domain is
partly comoving with the fluid, meaning that inside the star, both
edges of a domain are
comoving, the grid's velocity at each point being (linearly)
interpolated. The edge of the outer domain is at rest
($V_{grid}(r=R_{out})=0$), and the grid's velocity between star's surface and
this outer edge is interpolated at each point the same way. Typically,
two domains have been used for the star's interior and three for the
exterior, with about 65 points in each domain, which is considerably
lower than the number of points used in finite difference schemes. A
full run, with this multi-domain spectral methods, took about 20
minutes of CPU time on an Onyx Silicon Graphics workstation (with MIPS
R4400, 200 MHz processor), for $\sim 20000$ time steps.

\subsection{Tests}
\label{sa:test}

All equations have been checked using a Mathematica algebraic code.
More precisely, the Schwartz relation for equations (\ref{e:dmdr}) and 
(\ref{e:dmdt})
$$
{\partial^2 m \ov \partial r \partial t} - {\partial^2 m \ov \partial t
 \partial r} = 0
$$
has been computed and one then gets
$$
{\partial \varphi \ov \partial t} \times \l(\Box_{g_*} \varphi + q_\pi
 \alpha(\varphi)T_*\r) = {\partial \tilde{E} \ov \partial t} + {1\ov r^2} 
{\partial \ov \partial r} \bigl(r^2(\tilde{E}+\tilde{p})V\bigr)
- \Bigl(\text{ right-hand side of eq.~(\ref{e:dedt})}\Bigr)
$$
which is consistent with (\ref{e:boxfi}) and (\ref{e:dedt}).
Finally, setting $\varphi=0$ gives the equations of general relativity
as described in \cite{GOU91}.

For static configurations, the following test was performed: considering the
same equation
of state as in \cite{DEF93} (EOS1 in this work), the same coupling
function ($a(\varphi) = \exp(-3\varphi^2)$) 
and asymptotic scalar field value ($\varphi_0=0.0043$), we obtained
the same dependence for the effective scalar coupling constant (the
ratio between the scalar and the gravitational energies) on the star's
baryonic mass (Fig.2 of \cite{DEF93}). We also observed
an increase of the maximal gravitational mass of neutron stars, when
taking into account the scalar field, as it has been showed in Fig.~1
of \cite{DEF93}.

The subroutine solving the wave equation has been checked by taking
analytical solutions of simpler wave equations 
\be
{\partial ^2 \varphi \ov \partial t^2} = W(t,r)^2 \Delta \varphi
\ee
where $W(t,r)$ is a given function, and verifying that the
discrepancy between numerical solution and the analytical one goes
down as the square of the integration time step (second order
scheme). For example, 
$$
\varphi(t,r)={\tanh(r)\ov r}\ln(t+2)
$$
which is solution for $W(t,r)^2=\cosh(r)^2/2(2+t)^2\ln(t+2)$, was numerically
obtained at $10^{-6}$ relative accuracy, with a time step which is
 $dt^J=1/100$ of grid's radius, after $1000$ iterations.
The boundary condition has been checked by looking for the
remaining energy of the wave in the grid, after the wave was supposed
to be ``gone out''. With the same time-step as above, and for the
usual wave equation ($\Box \varphi=0$), after $500$
integrations, there remained $4.9\times 10^{-8}$ of the initial
energy, this amount going down as $(dt^J)^4$, for the energy is the
``square'' of the amplitude. As it has been seen in previous sections,
a numerical grid partly comoving with the fluid was used. For this
purpose, the wave equation have been adapted to such a grid and
tested.

Finally, during dynamical evolution, a good test of overall accuracy
was made by comparing the energy density $\tilde{e}$ given by the
baryonic density through the equation of state, to that deduced from
 $\tilde{E}$ (which is an evolved quantity). The same is possible for
 $H$. On a whole run this error always stayed $< \text{few} \times
10^{-3}$. The conservation of the baryonic number was verified with
a relative accuracy better than $10^{-5}$.

\newpage

\begin{figure}
\centerline{\epsfig{figure=db1030.ps,height=10cm,width=10cm,angle=270}}
\caption[]{\label{f:db1030}
Same as Fig.~\ref{f:db0m4} but
 $\beta_0$ varies from $10$ (upper curve, ---) to $\beta_0=30$
(lower curve, $-\cdot-$).}
\end{figure}

\begin{figure}
\centerline{\epsfig{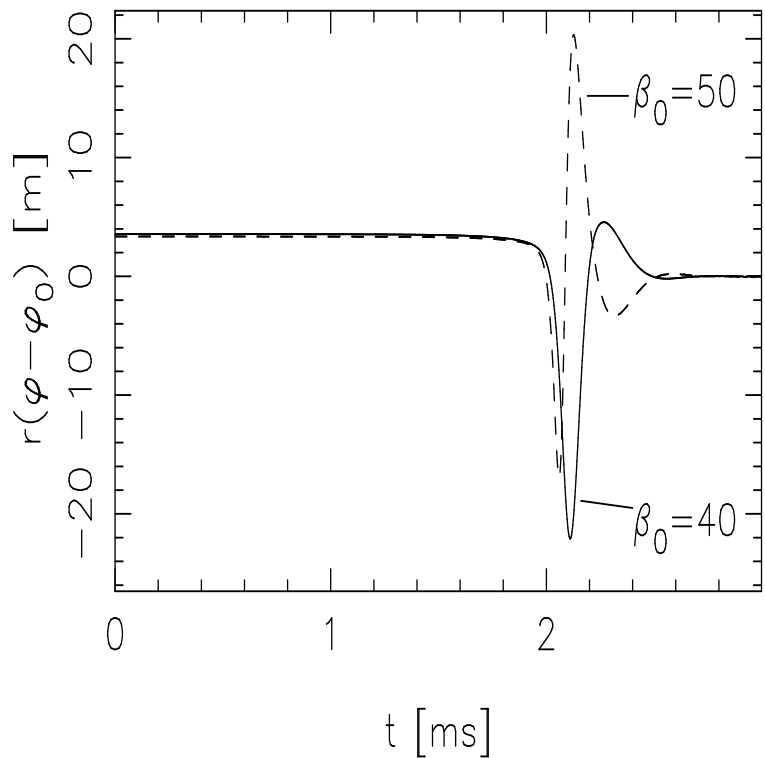}}
\caption[]{\label{f:db4050}
Same as Fig.~\ref{f:db0m4} but
 $\beta_0=40$ (---) and $\beta_0=50\ (---)$.}
\end{figure}

\begin{figure}
\centerline{\epsfig{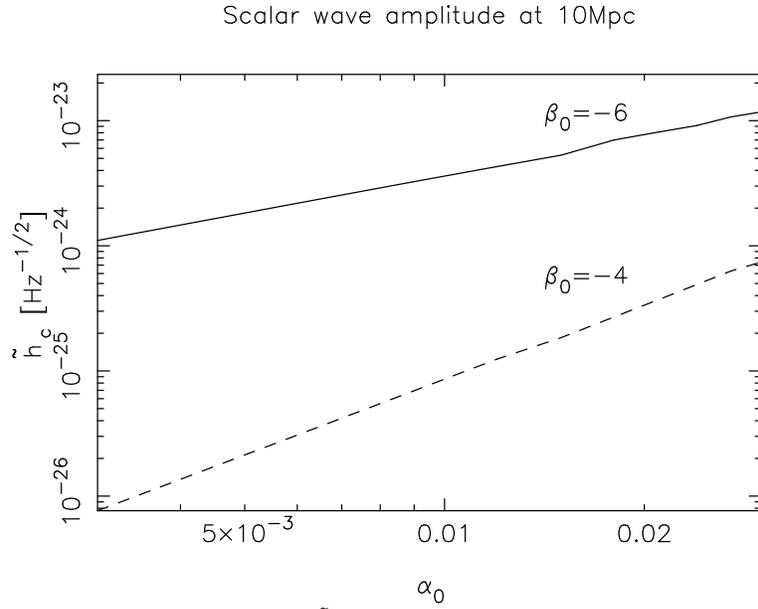}}
\caption[]{\label{f:dfidal}
Amplitude of the scalar gravitational wave $\tilde{h}_c$ (see
eq.~(\ref{e:hc})), at 10 Mpc, as a function of the coupling
coefficient $\alpha_0$, for two different values of $\beta_0$.}
\end{figure}

\begin{figure}
\centerline{\epsfig{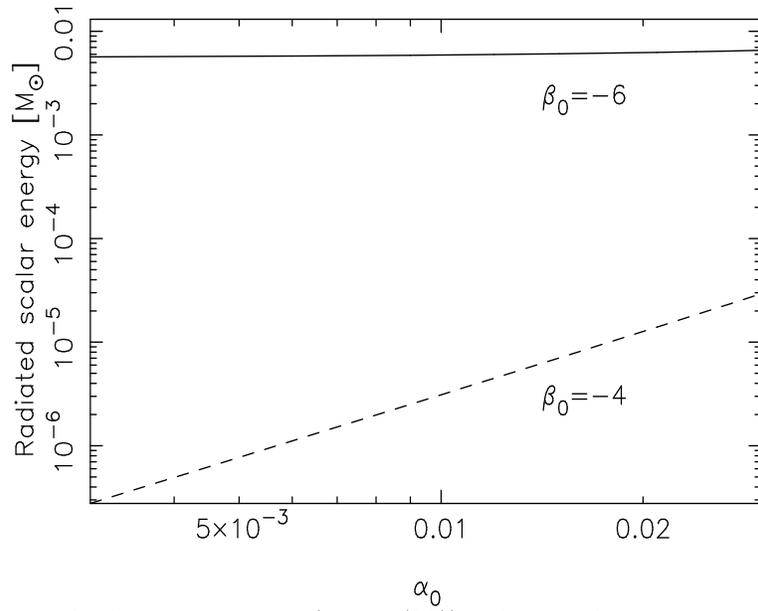}}
\caption[]{\label{f:erydal}
Radiated scalar gravitational energy $E_{scal}$ (see
eq.~(\ref{e:eray})) emitted during a collapse, as a function of the coupling
coefficient $\alpha_0$, for two different values of $\beta_0$.} 
\end{figure}

\begin{figure}
\centerline{\epsfig{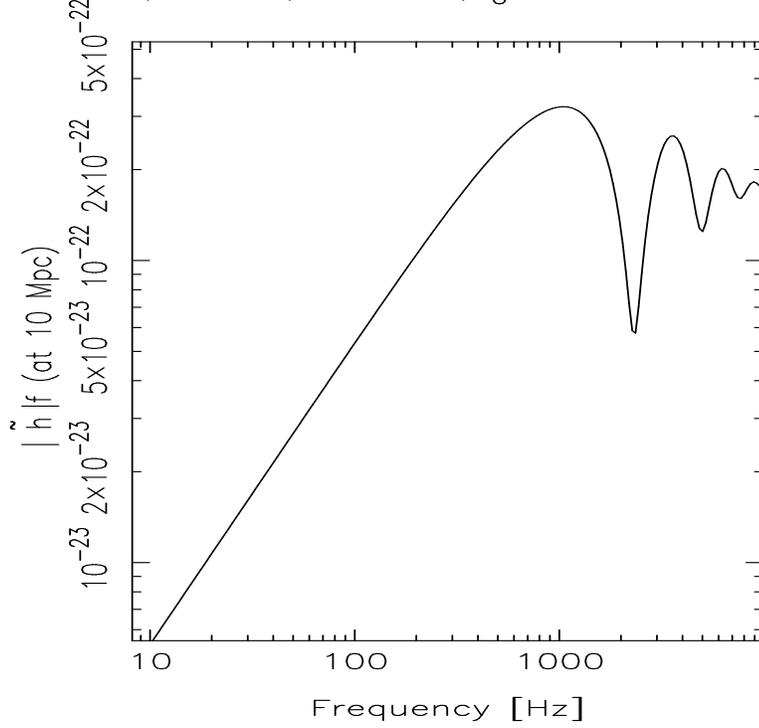}}
\caption[]{\label{f:fou256}
Same as Fig.~\ref{f:fou356}, but for collapse D.}
\end{figure}

\begin{figure}
\centerline{\epsfig{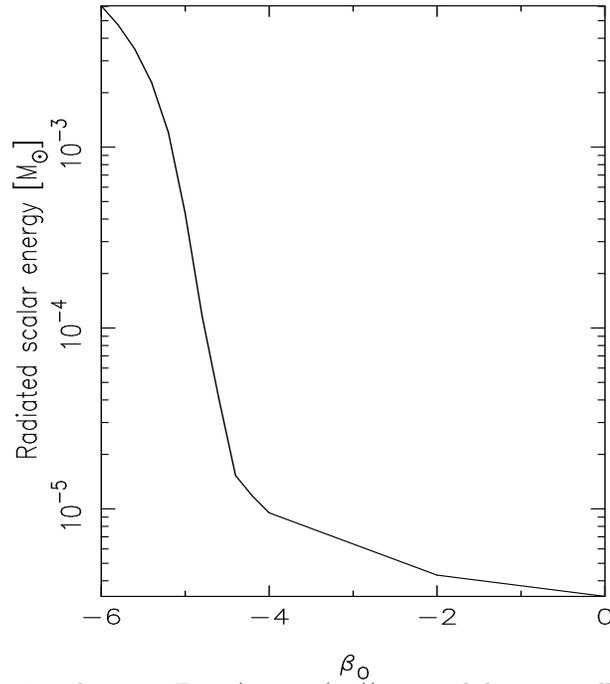}}
\caption[]{\label{f:eray1}
Radiated scalar gravitational energy $E_{scal}$ (see
eq.~(\ref{e:eray})) emitted during a collapse , as a function of the coupling
coefficient $\beta_0$, between $-6$ and $0$. For each $\beta_0$,
$\alpha_0$ has the 
maximal value, imposed by solar-system experiments (see
eq.~(\ref{e:alfamax})).}
\end{figure}

\begin{figure}
\centerline{\epsfig{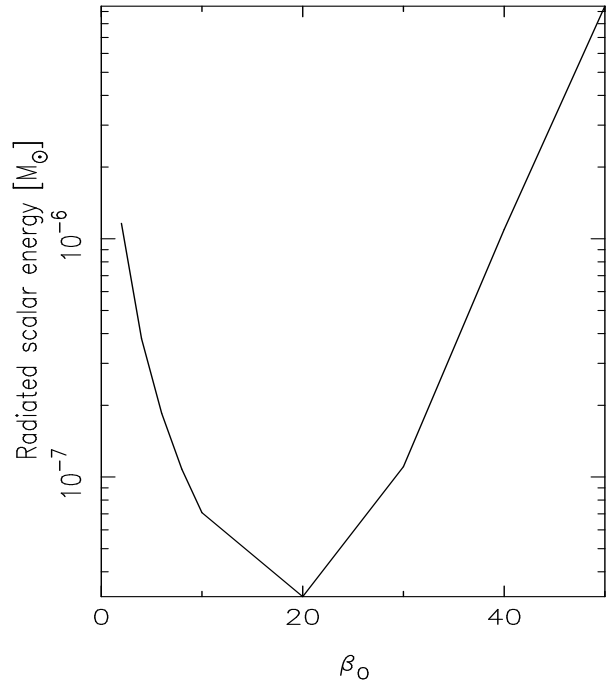}}
\caption[]{\label{f:eray2}
Same as Fig.~\ref{f:eray1} but $\beta_0$ varies from $0$ to $50$.}
\end{figure}

\begin{figure}
\centerline{\epsfig{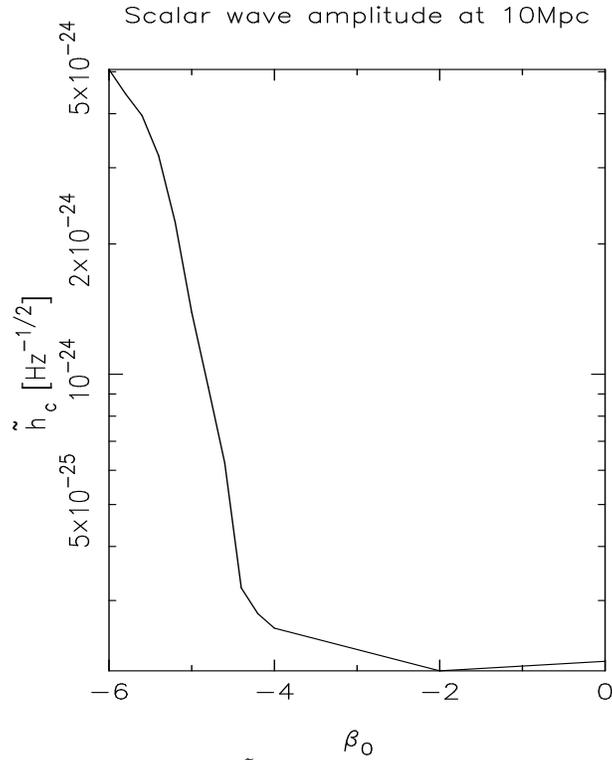}}
\caption[]{\label{f:deltfi1}
Amplitude of the scalar gravitational wave $\tilde{h}_c$ (see
eq.~(\ref{e:hc})), at 10 Mpc, as a function of the coupling
coefficient $\beta_0$, between $-6$ and $0$.  For each $\beta_0$,
$\alpha_0$ has the 
maximal value, imposed by solar-system experiments (see
eq.~(\ref{e:alfamax})).}
\end{figure}

\begin{figure}
\centerline{\epsfig{figure=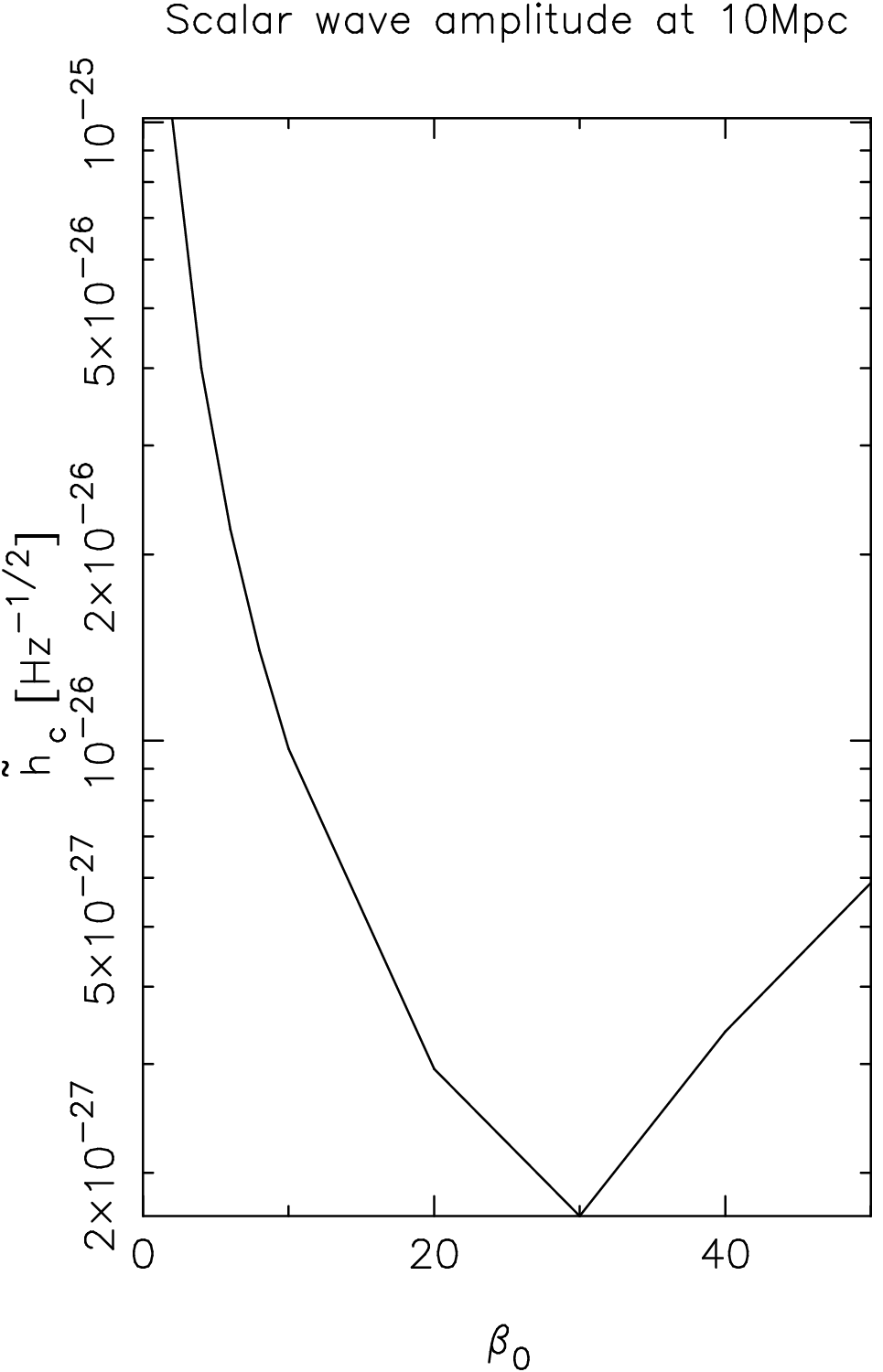,height=12cm,width=8cm,angle=0}}
\caption[]{\label{f:deltfi2}
Same as Fig.~\ref{f:deltfi1}  but $\beta_0$ varies from $0$ to $50$.}
\end{figure}

\newpage

\begin{table}
\caption{\label{t:coll}
Initial condition parameters of the collapses presented in this
paper. The equations of state (EOS) are described in
sec.~\ref{s:stat}, $\varphi_0$ is the asymptotic scalar field value
(given by cosmological evolution), $\alpha_0$ and $\beta_0$ are the
coupling function parameters (\ref{e:cfunc}), $R_{star}$ denotes
star's radius, $\tilde{n}_B(r=0)$is the central baryon density (in
units of nuclear density, $1\ n_{nuc}=10^{44}\ \text{m}^{-3}$), $M_G$
is the $g^*_{\mu\nu}$-frame ADM mass, $M_B$ the baryonic one and
$\omega$ the scalar charge.}
\begin{tabular}{c|c|c|c|c|c|c|c|c|c}
Collapse & EOS & $\varphi_0$ & $\alpha_0$ & $\beta_0$ & $R_{star}$ &
 $\tilde{n}_B(r=0)$ & $M_G$ & $M_B$ & $\omega$ \\
&&&&& [km] & [$n_{nuc}$] & [$\ M_\odot$] & [$\ M_\odot$] & [$\
 M_\odot$] \\
\hline
A & 1 & $10^{-5}$ & $5\times 10^{-5}$ & $-5$ & $11.2$  &
$10.4$ & $1.97$ & $2.26$ & $0.204$ \\

B & 1 & $10^{-5}$ & $2.5\times 10^{-2}$ & $-5$ & $11.8$ &
 $10.4$ & $2.07$ & $2.41$ & $0.484$ \\

C & 2 & $10^{-5}$ & $5\times 10^{-5}$ & $-5$ & $21.5$  & $4$ &
 $3.31$ & $3.68$ & $0.921$ \\

D & 2 & $10^{-5}$ & $2.5\times 10^{-2}$ & $-5$ & $22.2$ & $4$
& $3.41$ & $3.82$ & $1.16$ \\
\end{tabular}
\end{table}


\begin{references}
\bibitem{DEF92} T.~Damour and G.~Esposito-Far\`ese, Class. Quantum Grav.
{\bf 9}, 2093 (1992)
\bibitem{FIE56} M. Fierz, Helv. Phys. Acta {\bf 29}, 128 (1956) 
\bibitem{JOR59} P. Jordan, Z. Phys. {\bf 157}, 112 (1959)
\bibitem{BD61} C. Brans and R.H. Dicke, Phys. rev. {\bf 124}, 925 (1961)
\bibitem{BER68} P.G. Bergmann, Int. J. Theor. Phys. {\bf 1}, 25 (1968)
\bibitem{NOR70} K. Nordtvedt, Astrophys. J. {\bf 161}, 1059 (1970)
\bibitem{WAG70} R.V. Wagoner, Phys. Rev. D {\bf 1}, 3209 (1970)
\bibitem{CFM85} C.G. Callan, D. Friedman, E.J. Martinec and M.J. Perry,
Nucl. Phys. B {\bf 262}, 593 (1985)
\bibitem{DP94} T. Damour and A.M. Polyakov, Nucl. Phys. B {\bf 423}, 532 (1994)
\bibitem{SA90} P.J Steinhardt and F.S. Accetta, Phys. Rev. Lett. {\bf 64},
2740 (1990)
\bibitem{DEF93} T. Damour and G. Esposito-Far\`ese, Phys. Rev. Lett. {\bf 70},
2220 (1993)
\bibitem{DEF96} T. Damour and G. Esposito-Far\`ese, Phys. Rev. D {\bf 54},
1474 (1996)
\bibitem{VIRGO} C. Bradaschia, {\em et al.}, Science {\bf 256}, 325 (1992)
\bibitem{LIGO} K.S. Thorne, in {\em Proceedings of the International
 Conference on Particle Physics, Astrophysics and Cosmology} (1996), p41
\bibitem{SNN94} M. Shibata, K. Nakao and T. Nakamura, Phys Rev. D {\bf 50},
6058 (1994)
\bibitem{SST95} M.A. Scheel, S.L. Shapiro and S.A. Teukolsky, Phys. Rev. D
{\bf 51}, 4236 (1995)
\bibitem{HCN97} T. Harada, T. Chiba, K. Nakao and T. Nakamura, Phys. Rev. D
{\bf 55}, 2024 (1997)
\bibitem{GOU91} E. Gourgoulhon, Astron. Astrophys. {\bf 252}, 651 (1991)
\bibitem{ADM62} R. Arnowitt, S. Deser and C.W. Misner, in {\em Gravitation}
Witten L. (ed.) Wiley, New York (1962)
\bibitem{PIR83} T. Piran, in {\em Rayonnement gravitationnel -- Les Houches
 1982} N. Deruelle and T. Piran (eds.) North Holland, Amsterdam (1983)
\bibitem{DAI85} J. Diaz-Alonso and J.M. Iba\~nez-Cabanell,
Astrophys. J. {\bf 291}, 308 (1985)
\bibitem{SBG94} M. Salgado, S. Bonazzola, E. Gourgoulhon and
P. Haensel, Astron. Astrophys. {\bf 291}, 155 (1994)
\bibitem{CEF90} R. Coquereaux and G. Esposito-Far\`ese, Ann. Inst. H.
Poincar\'e {\bf 52}, 113 (1990)
\bibitem{BOM86} S. Bonazzola and J.A. Marck, Astron. Astrophys. {\bf 164},
300 (1986)
\bibitem{WK97} R.V. Wagoner and D. Kalligas, in {\em Relativistic
Gravitation and Gravitational Radiation } J. A. Marck and J. P. Lasota
(eds.) Cambridge University Press (1997)
\bibitem{HAR97} T. Harada, Prog. Theor. Phys. (preprint {\bf 98}, 359 (1997)
\bibitem{VIR2} B. Caron, {\em et al.}, in {\em Gravitational Waves:
Sources and Detectors} I. Ciufolini and F. Fidecaro (eds.) World
Scientific Publishing (1997) 
\bibitem{BOM93} S. Bonazzola and J.A. Marck, Astron. Astrophys. {\bf
267}, 623 (1993)
\bibitem{ZMU97} T. Zwerger and E. M\"uller, Astron. Astrophys. {\bf
320}, 209 (1997)
\bibitem{BOM90} S. Bonazzola and J.A. Marck, J. Comp. Phys. {\bf 87},
201 (1990)
\bibitem{BGM97} S. Bonazzola, E. Gourgoulhon and J.A. Marck, in {\em 
Relativistic Gravitation and Gravitational Radiation } J. A. Marck and
J. P. Lasota (eds.) Cambridge University Press (1997)
\end{references}
\end{document}